\newcounter{myctr}

\documentclass{ws-acs}
\usepackage{url}
\usepackage{caption}
\usepackage{array}

\renewcommand{\theenumi}{\roman{enumi}}%

\begin{document}

\makeatletter
\def\@biblabel#1{[#1]}
\makeatother

\markboth{Evgeny Ivanko}{Should Evolution Necessarily be {\it Ego}lution?}

%
%

\title{SHOULD EVOLUTION NECESSARILY BE {\it EGO}LUTION?
}

\author{\footnotesize EVGENY IVANKO
\ \\
\ \\
}

\address{Institute of Mathematics and Mechanics, Ural Branch, Russian Academy of Sciences\\
S. Kovalevskoi 16, Ekaterinburg, 620990,
Russia
\\
\ \\
Ural Federal University\\
Mira 19, Ekaterinburg, 620002, Russia\\
\ \\
evgeny.ivanko@gmail.com}

\maketitle

\begin{history}
\end{history}

\begin{abstract}
In the article I study the evolutionary adaptivity of two simple population models, based on either altruistic or egoistic law of energy exchange. The computational experiments show the convincing advantage of the altruists, which brings us to a small discussion about genetic algorithms and extraterrestrial life.
\end{abstract}


\section{Introduction}

The origins and potential of self-developing complex processes is one of the most challenging topics today and will become one of the most important branches of science tomorrow. As a part of Earth's biosphere we are familiar in practice with only one such a process -- evolution. Although the evolution of biosphere on Earth may be studied in detail, there are no other known examples of natural self-developing complexity, which leaves us with our imagination and artificial life models on the way to the essence of autonomous systems' organization and complication. One of the recently proposed instruments to cope with the lack of heterogeneous examples is \textit{radical reimplementation} \cite{radical}, where experience in some vague subject is variegated by achieving the key aims using the most unusual means. This article is focused on the radical reimplementation of selfish motive in evolution.

Evolution is usually considered as a process driven by individual reproduction gain. Even altruistic phenomena like reciprocal altruism \cite{recip2,recip1} or eusociality \cite{altru4} with its mechanisms of inclusive fitness \cite{incl1} often thought to be evolved on the basis of ``selfish gene'' ground. There are many works devoted to the appearance of altruistic phenomena in egoistic world (see, e.g. \cite{altru6,altru11,altru18,altru19,altru1,altru20,altru16,altru22}). But I am familiar with only one, where altruism is considered as a preliminary inherent property of artificial life system, and the resulting model competes with the egoistic-based model in  efficiency \cite{altru9}. The current paper continues this competition: altruistic- and selfish-based models with simple homogeneous population structure are exposed to the  evolution process in a changing environment with the purpose to explore which strategy provides better adaptation and survival.
Compared with the well-studied approach of \textit{evolutionary games} \cite{games5,games12,games1,games15}, in this study both altruistic and selfish strategies struggle not against each other, but against natural obstacles lurking for a population in mutation-selection process. Experiments with  ``homogeneous'' populations whose organisms adhere to one single strategy allow to estimate the adaptation flexibility of each strategy and to come closer to the answer to the question ``is egoism a necessary attribute of evolution.''


The model populations consist of moving colored circles, where circle's radius is energy and RGB-color is 3-genes genome (Fig. 1). Each time two circles meet they exchange energy in either altruistic (the larger one gives a half of its energy to the smaller one) or selfish (the larger strips a half of the smaller's energy) way. If an organism becomes large enough, it disappears giving birth to 2 new organisms. The probability of such replication depends on  the organism's fitness -- the correspondence between the current ``environmental conditions'' (which may be visualized as the color of the background field and may be thought of as the current  ``target genome'') and the organism's genome. Successive changes of the background color forces the populations to move through the ``labyrinth of adaptation.'' In classification of intriguing work \cite{org}, the altruistic populations show high cooperation and low conflict, which corresponds to the type ``\textit{organism}''; in case of the selfish behavior, the elements of the populations show low cooperation and high conflict corresponding to a ``group of \textit{competitors}.''

The first question under experimental investigation is plain survival of the populations equipped with one of the two strategies in a rapidly and roughly changing environment (\textit{stress-test}). The populations that succeed to beget consequently all the 8 given target genomes without stumbling across genetic bottlenecks \cite{bottle2,bottle1} or stasis \cite{stasis} are then compared by  speed of adaptation. 

The second question is how long a population can stay in a constant environment (and thus ``overspecialize'') while preserving enough genetic variety to be able to survive and adapt when the environment finally radically changes (\textit{idyll-test}). The results of the conducted experiments suggest that the altruistic populations are much more flexible in the process of artificial evolution.

\section{Model description}


The population structure and evolution process models are rather simple and resemble bacterial life in a changing milieu.
\paragraph{Organisms.}
Each organism $O$ is represented by a circle, whose radius $rad(O)$  corresponds to``energy,'' and 3-component RGB-color $(r(O),g(O),b(O))\in\overline{1,255}^3$ corresponds to 3-gene  ``genome.'' The position of the center of each organism is given by the two coordinates $(x(O),y(O))$; discrete-time dynamics is determined by the constant speed vector $\mathrm{\bf v}(O)$ with the randomly chosen unchangeable direction $\alpha(O)$ and the length  equal to 2. Every \textit{step}, each organism is shifted according to the simple linear rule:
\begin{align}
x(O):=x(O)+|\mathrm{\bf v}(O)|cos(\alpha(O)),\nonumber\\
y(O):=y(O)+|\mathrm{\bf v}(O)|sin(\alpha(O)).
\end{align}
The speed-vector length 2 is just a practical compromise: its increase
 intensifies evolution speed (which is a plus as it allows to conduct more computational experiments), but also enhances the chance of contact inaccuracy (the situation where two small organisms moving discretely can just step over each other without contact).

\paragraph{Field.}

\begin{figure}
\begin{center}
\includegraphics[width=0.2\textwidth]{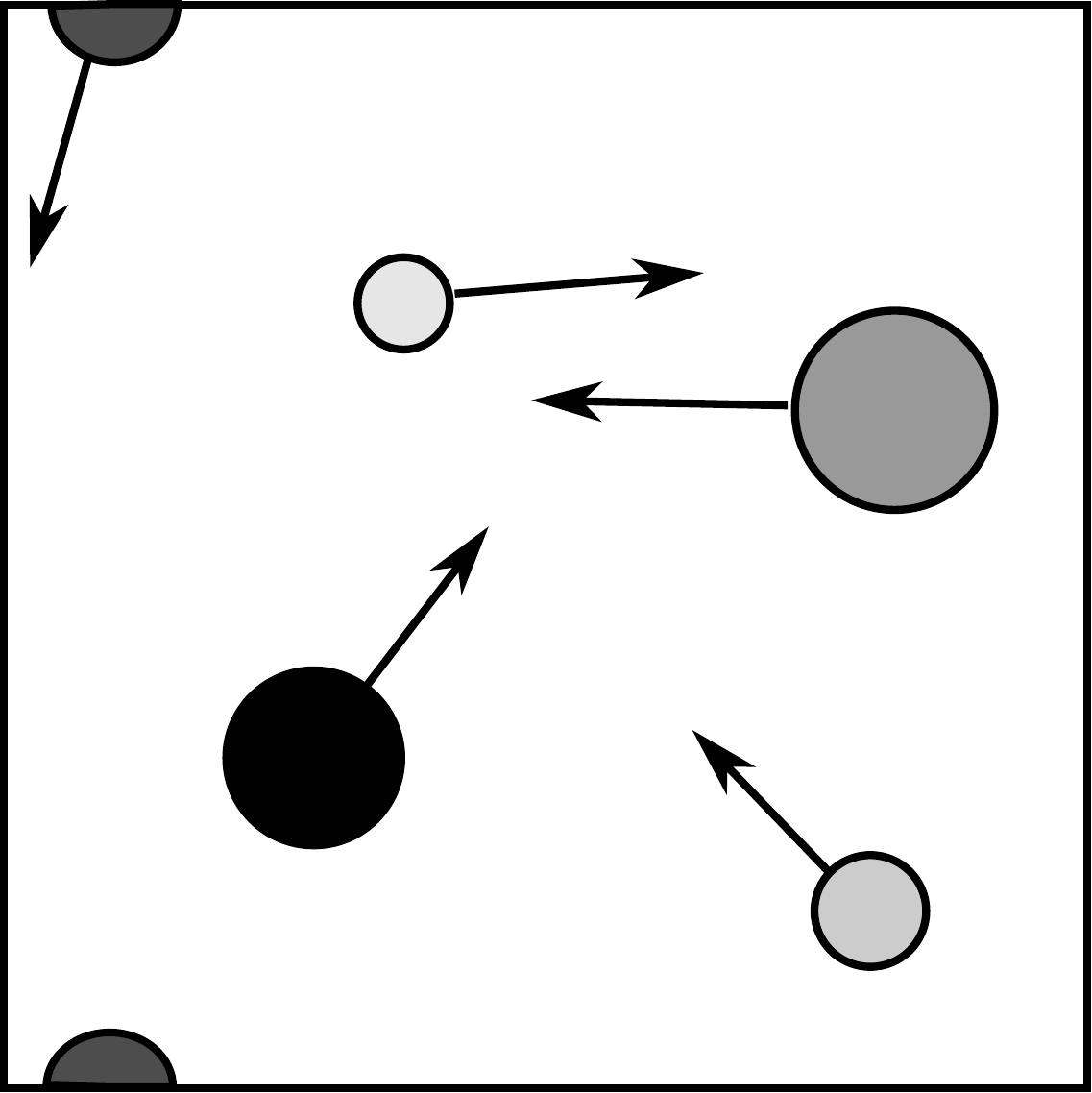}
\caption{Experiment field}
\label{fig:1}
\end{center}
\end{figure}

The organisms live inside a discrete plane square field $\overline{1,100}\times \overline{1,100}$ with rather common toric topology, which means that  if an organism moves behind a border it appears out of the opposite border with the same speed vector (see Figure \ref{fig:1}).

\paragraph{Population.}
Each population initially consists of 100 organisms, and this is the maximum allowed number of organisms in the population. There are three important parameters that characterize each population: 
\begin{itemize}
\item mature size ($ms$)  an organism that reaches this radius may divide in 2 newbie organisms (unless the population limit is already reached); 
\item die size ($ds$) is the radius falling below which an organism dies; 
\item newbie size ($ns$) is the radius of any newborn organism.
\end{itemize}
The initial conditions -- positions, radii and colors of the population's organisms -- are chosen uniformly: $(x,y)\in\overline{1,100}^2$, $rad\in\overline{ds,ms}$, $(r,g,b)\in\overline{1,255}^3$.

It is evident that any viable population has to receive more energy, than it spends. In the following experiments, the energy input is provided by two simple conventions:
\begin{itemize}
\item $ns>ds$ -- an evident survival constraint;
\item $ns\le ms\le 2ns$ emulates a reasonable energy input to the model (analog of plentiful food; large newbie size corresponds to the imaginary situation where an ancestor lets his two descendants participate in social life only when they are grown to some extent).
\end{itemize}



\paragraph{Contact.}

Generally, the organisms are transparent for each other except for the ``touch'' moment. Two organisms are considered to be in touch at the current step if:

1) the distance between their centers does not exceed the sum of their radii,
\begin{equation}
\label{touch}
\sqrt{(x(O_1)-x(O_2))^2+(y(O_1)-y(O_2))^2}\le rad(O_1)+rad(O_2);
\end{equation}

2) the condition \eqref{touch} was not true at the previous step.

\paragraph{Energy exchange.}

When two organisms $O_1$ and $O_2$ touch each other they exchange  energy in either \textit{altruistic} or \textit{selfish} way. 
In an \textit{altruistic} exchange, the \textit{larger} organism \textit{gives} a half of its energy  to the smaller one:
\begin{align}
\label{altruist1}
\mathrm{if}\ \ rad(O_1)&\ge rad(O_2)\ \mathrm{then}\nonumber  \\
& rad(O_2):=rad(O_2)+rad(O_1)/2,\nonumber\\
& rad(O_1):=rad(O_1)-rad(O_1)/2; 
\end{align}
in a \textit{selfish} exchange vice versa the \textit{larger} organism \textit{takes} a half of the energy of the smaller one:
\begin{align}
\label{altruist2}
\mathrm{if}\ \ rad(O_1)&> rad(O_2)\ \mathrm{then} \nonumber \\
& rad(O_1):=rad(O_1)+rad(O_2)/2,\nonumber\\
& rad(O_2):=rad(O_2)-rad(O_2)/2. 
\end{align}

\paragraph{Fission.}

If the size of an organism $O$ exceeds the mature size $ms$ and the number of the organisms in the population is less than 100, the organism gets a chance to split into two equivalent newbie organisms ($O$ is replaced with $\{O_1,O_2\}$).

The probability $p$ to split reflects the organism's ``fitness'' and is inversely related to the ``distance'' between the  organism's genome $(r(O),g(O),b(O))$ and the current target genome $(r_0,g_0,b_0)$ (which may be considered as the best possible genome for the ``current environment,'' see the next subsection for the details):

\begin{equation}
\label{rep_prob}
p=1-\sqrt{\frac{((r(O)-r_o)^2+(g(O)-g_o)^2+(b(O)-b_o)^2)}{3\cdot255^2}}.
\end{equation}
The newborn organisms inherit some of the parent's properties:
\begin{align*}
x(O_1)=x(O_2)=x(O),\quad
y(O_1)=y(O_2)=y(O),\quad
|\mathrm{\bf v}(O_1)|=|\mathrm{\bf v}(O_2)|=2.
\end{align*}
The angles $\alpha(O_1)$ and $\alpha(O_2)$ are selected uniformly from $[0,2\pi)$, the size is predefined at $rad(O_1)=rad(O_2)=ns$. 
 The genomes of the newborn organisms are formed by a random mutation:
\begin{align}
r(O_i)=\min\{\max\{0,r(O)+\delta^i_r\},255\},\nonumber\\
g(O_i)=\min\{\max\{0,g(O)+\delta^i_g\},255\},\nonumber\\
b(O_i)=\min\{\max\{0,b(O)+\delta^i_b\},255\},
\end{align}
where $i\in \{1,2\}$ and $\delta^i_r,\delta^i_g,\delta^i_b$ are uniformly selected from the integer interval $\overline{-5,5}$.

\begin{figure}
\centering
  \includegraphics[width=0.4\textwidth]{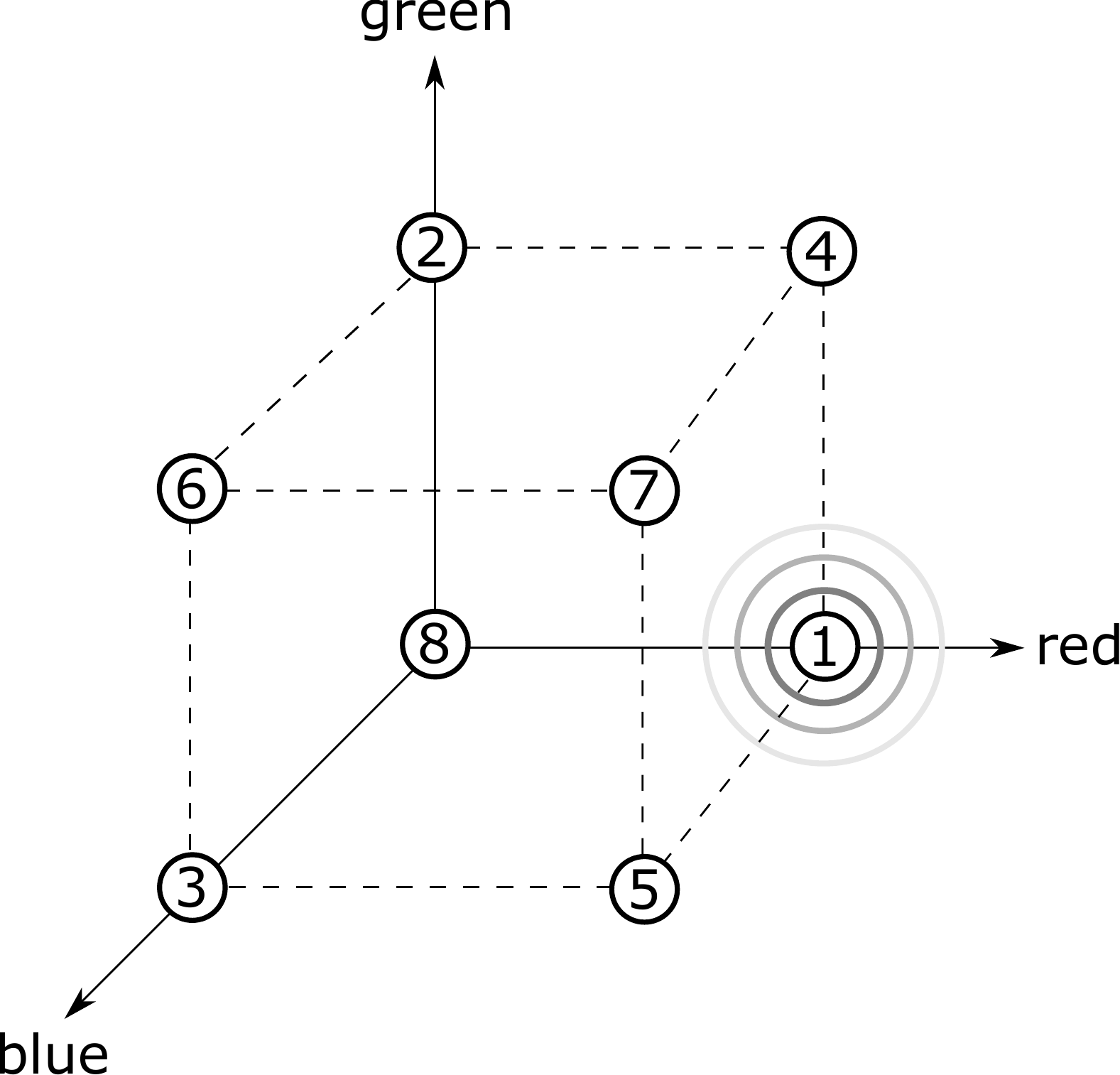}
\caption{RGB-cube, the sequence of the target genomes. The first position has to be reached from a random state, which is indicated by concentric circles
}
\label{fig:2}       
\end{figure}

\paragraph{Changing environment.}
One of the main purposes of this article is to see how successful are the altruistic and selfish strategies in a radically changing environment. The changes in the population's environment here are modeled by subsequent changes in the replication probability function \eqref{rep_prob}, each of which consists in a modification of the target genome $(r_0,g_0,b_0)$. All the populations have to follow through the same sequence of target genomes:
\begin{align}
\big((255,0,0),(0,255,0)&,(0,0,255),(255,255,0),\nonumber\\&(255,0,255),(0,255,255),(0,0,0),(255,255,255)\big),
\end{align}
which may be visualized as the vertices of the RGB-color cube (see Figure \ref{fig:2}). The environment (target genome) changes right after  the population ``adapts to the current environment.'' In the following experiments it means that the fitness function \eqref{rep_prob} switches to the next  target genome as soon as the population produces an organism, whose genome is \textit{equal} to the current target genome.

\paragraph{Success criterion.}

All the studied populations were subjected to the following two main tests, revealing the ability to survive in unpredictable process of evolution:

\begin{enumerate}
\item \textit{stress-test}, how fast, manoeuvrable and stable a population is in a stressful environment; formally we will measure it by the number of the steps needed to reach the 8-th target genome and by the minimum number of different genomes in the population during its evolution (genetic diversity);
\item \textit{idyll-test}, how long a population is able to continue the ``overspecialization'' in constant conditions after the first target genome is reached while preserving the ability (i.e., a sufficient diversity) to turn the evolution toward a radically different target genome and successfully beget it.
\end{enumerate}

It is impossible to follow the destiny of a population forever; fortunately it is not necessary since the population discloses its potential relatively fast. The following are the empiric rules that were used in the computational experiments to cull the ``unpromising'' populations.

\begin{itemize}
\item {\bf Rule $\mathbf{10^6}$.} A population must reach the last 8-th target genome in $10^6$ steps. On the one hand, this is long enough for many populations to succeed, on the other hand, the experiments show that those that can not do it have a little chance to succeed even in $10^8$ steps.
\item {\bf Rule $\mathbf{10^4}$.} A population must reach its first target genome in $10^4$ steps. This is an instrument to cull the populations that evolve too slow (since it is a relatively easy task to get to the first target genome starting with the high genetic diversity of the initial uniform state). Practice shows that if a population cannot meet Rule $10^4$, it  has a disappearing chance to meet the above Rule $10^6$.
\item {\bf Rule 3.} A population must contain more than 3 organisms at each step. This is an  instrument to cull the populations that succumb to degeneration. In practice, if the number of organisms gets down to 3 or less, it is highly probable that the abundance would not recover; such populations have a poor chance to meet the first Rule $10^6$.
\end{itemize}

\section{Experiments}

\paragraph{Stress-test.}
In the first test each population was adapting to the environment, which changed each time a population succeeded in begetting an ``ideal'' genome for the current environment. In accordance with the survival and energy delivery constraints from  Section 2.3, the following parameters were used to produce the experiment populations:
\begin{align}
\label{constraints}
ms, ds, ns\in\overline{0,10};\ \ \
ns>ds;\ \ \ ns<ms\le 2ns.
\end{align}
For each tuple of the parameters' values $(ms,ds,ns)$, 100 independent computational experiments were conducted. During each experiment, a population consequently pursued 8 target genomes starting from a random genetic state. In case at least one of the rules (Rule $10^6$, Rule $10^4$ or Rule 3) was broken, the experiment was terminated, the abort reason was recorded and the computation proceeded to the next experiment starting with a new random state.

\begin{figure}
\centering
\begin{tabular}{>{\centering\arraybackslash}m{0.2cm}>{\centering\arraybackslash}m{5cm}>{\centering\arraybackslash}m{5cm}}
& Selfish & Altruistic\\
1&\includegraphics[width=0.29\textwidth]{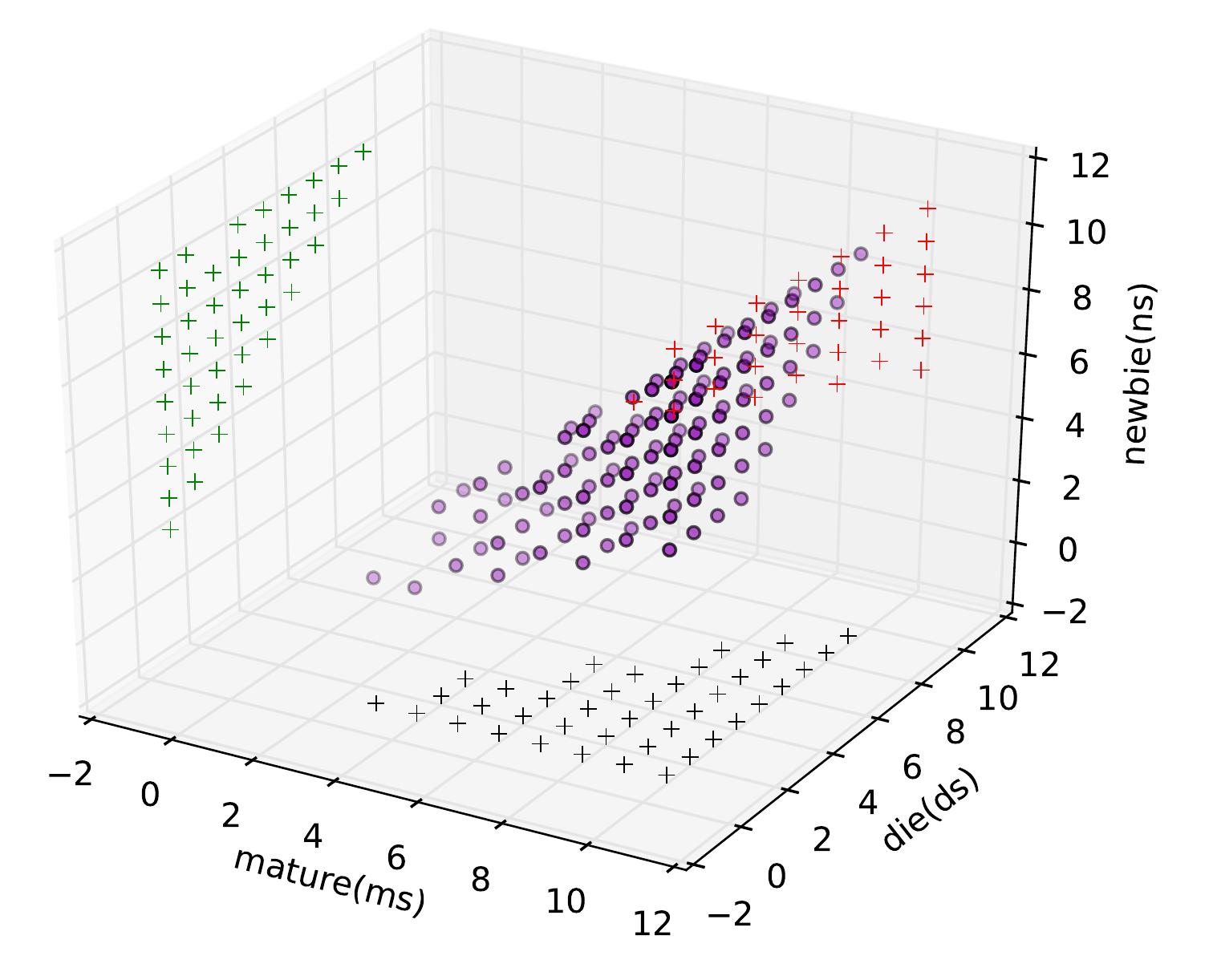}& \includegraphics[width=0.29\textwidth]{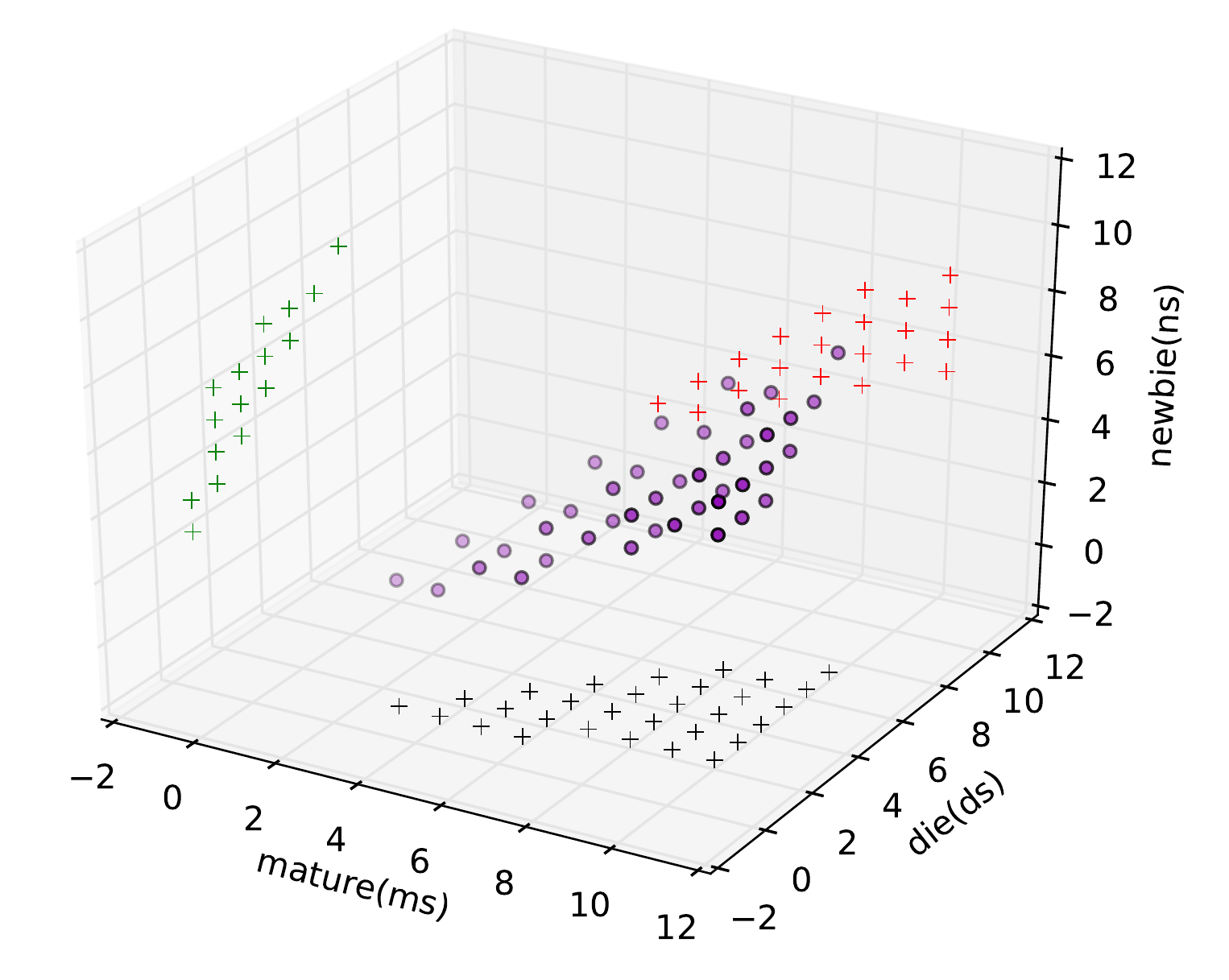} \\
2&\includegraphics[width=0.29\textwidth]{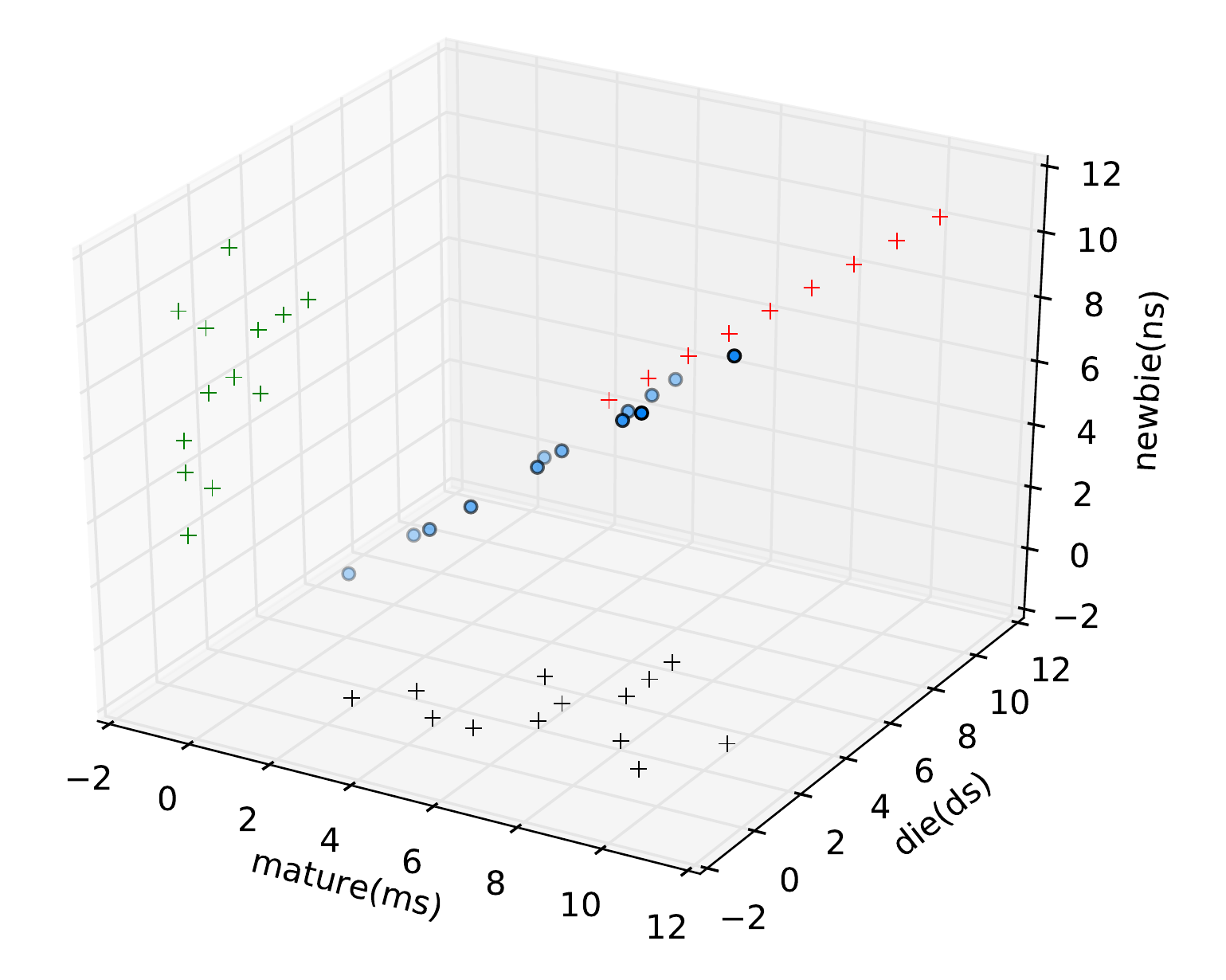}& \includegraphics[width=0.29\textwidth]{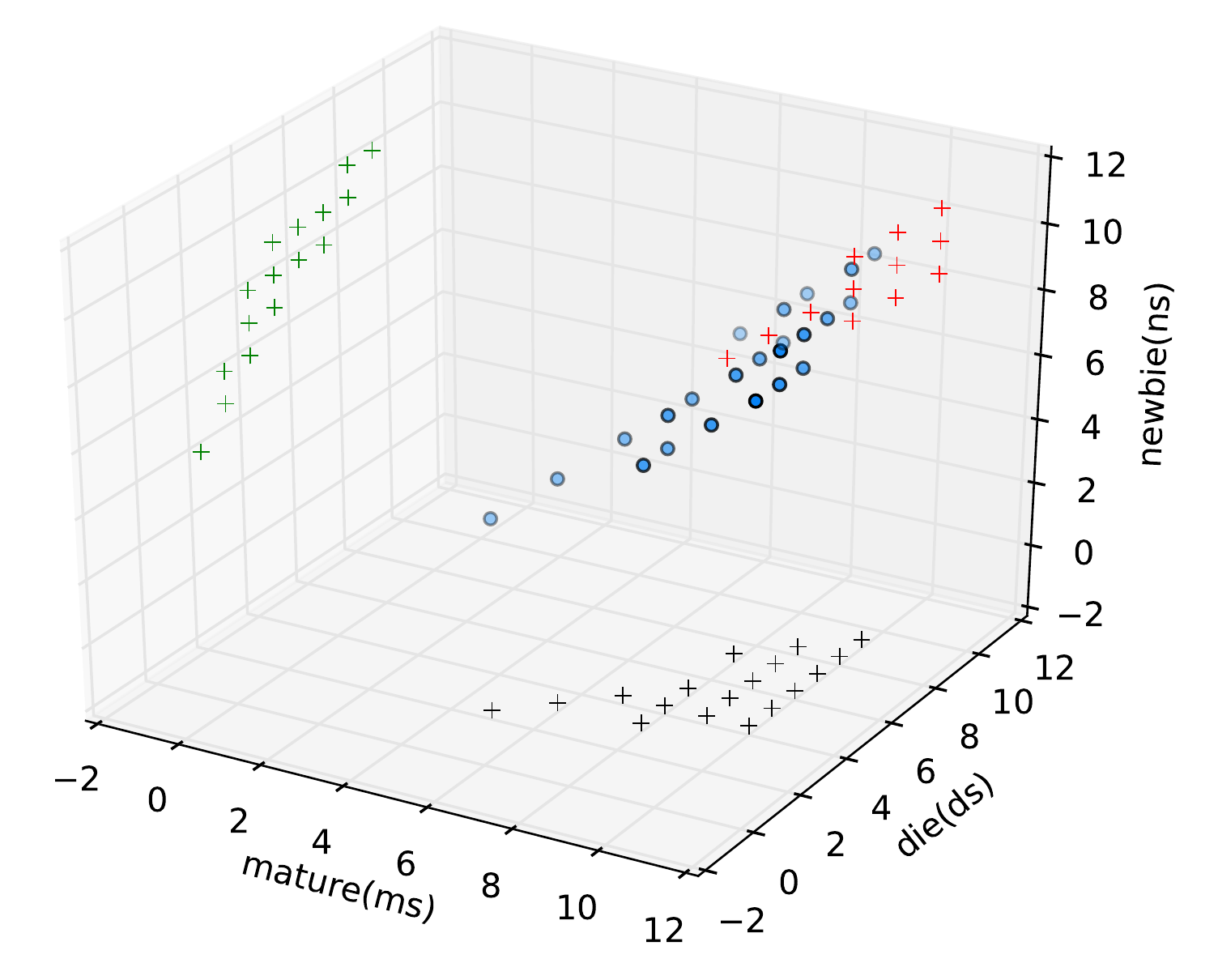} \\
3&\includegraphics[width=0.29\textwidth]{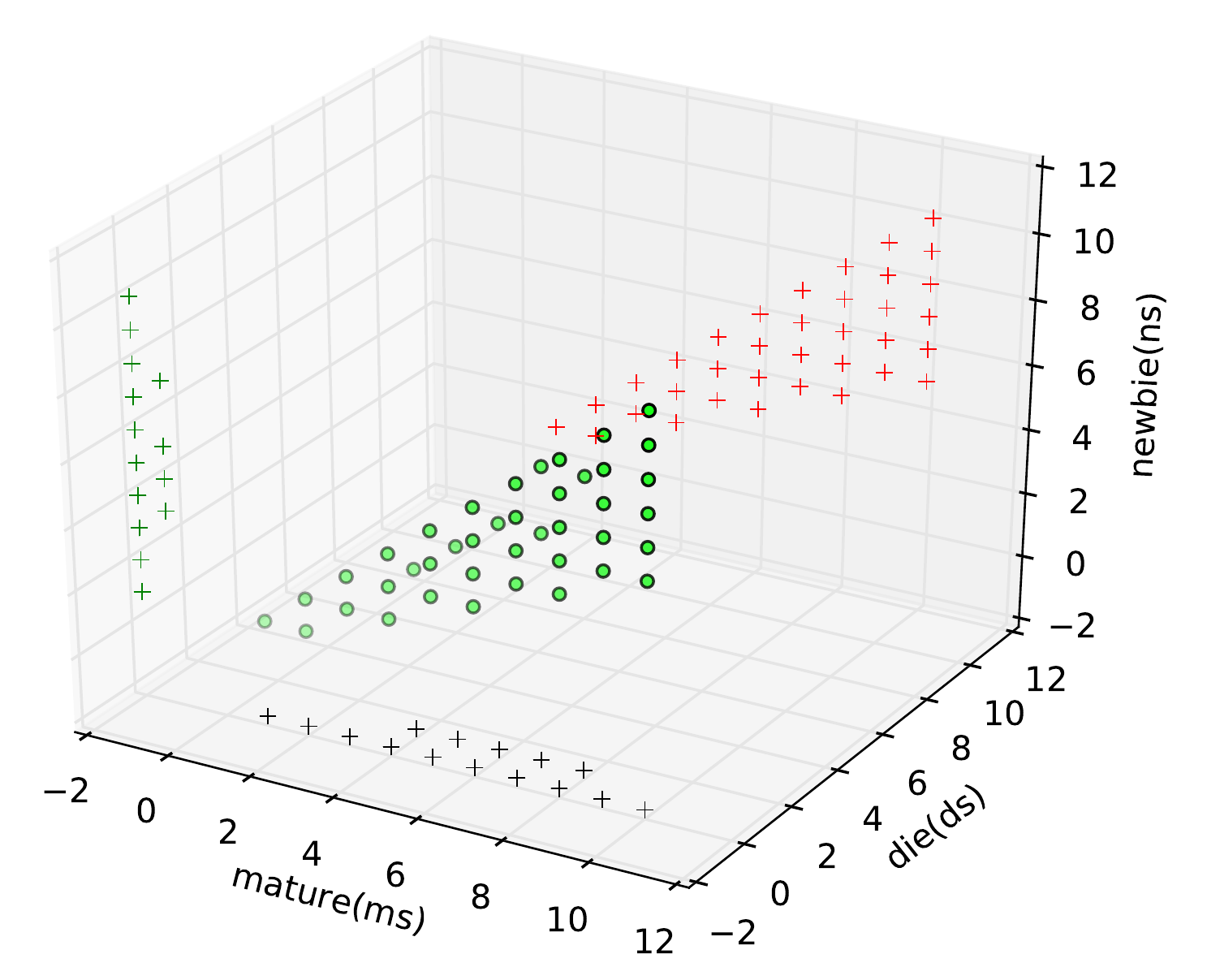}& \includegraphics[width=0.29\textwidth]{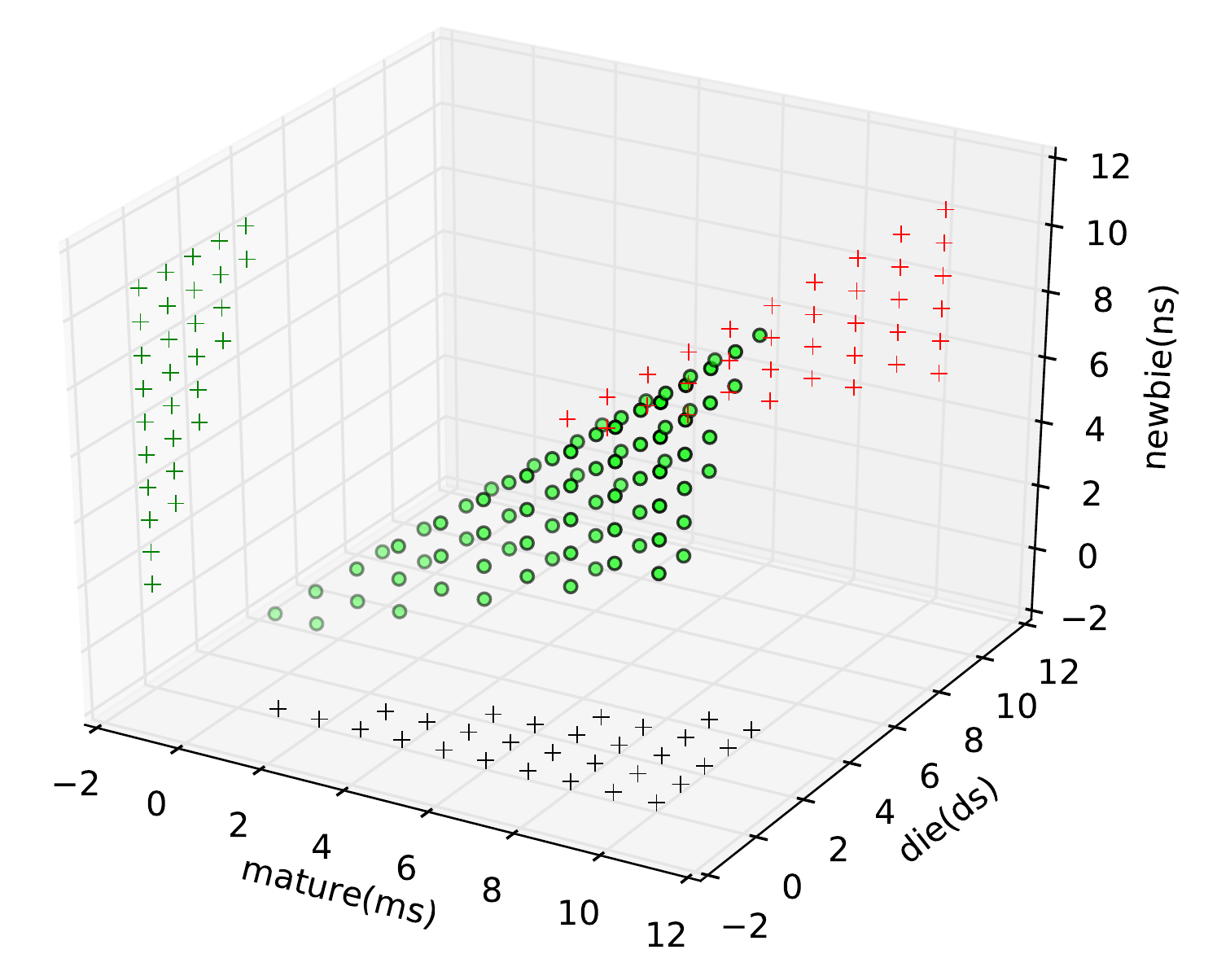} \\
4&\includegraphics[width=0.29\textwidth]{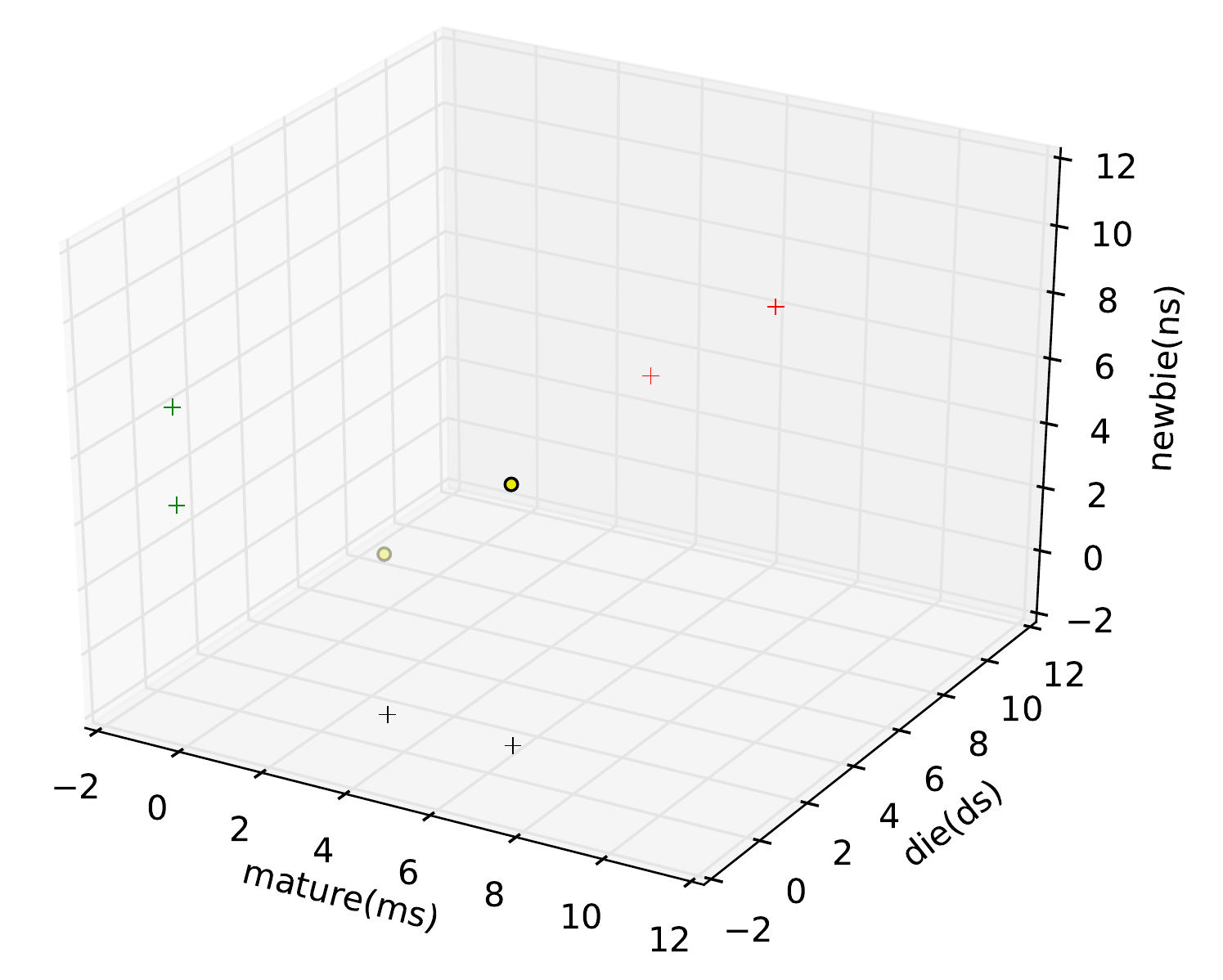}& \includegraphics[width=0.29\textwidth]{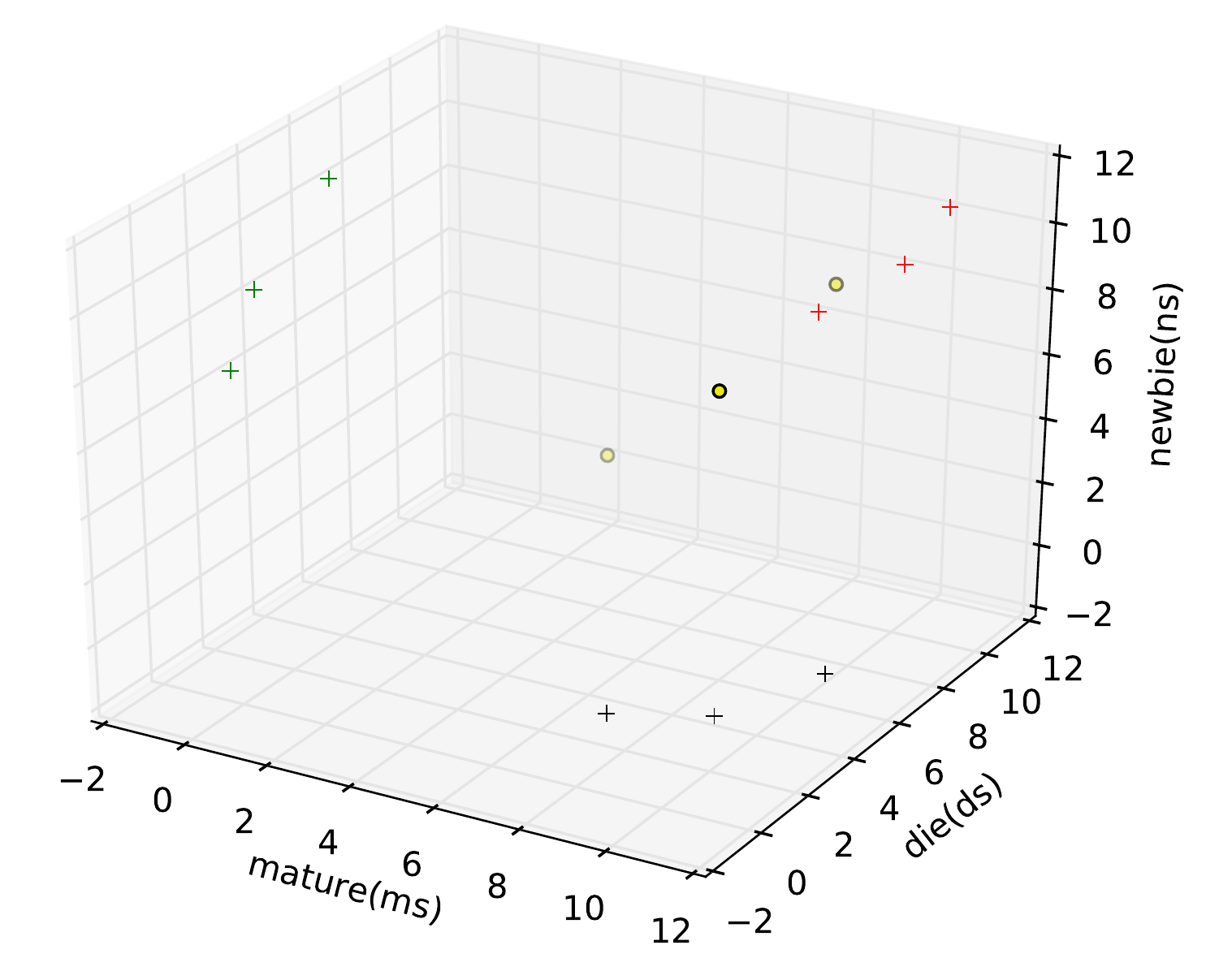} \\
5& - & - \\
6& - & \includegraphics[width=0.29\textwidth]{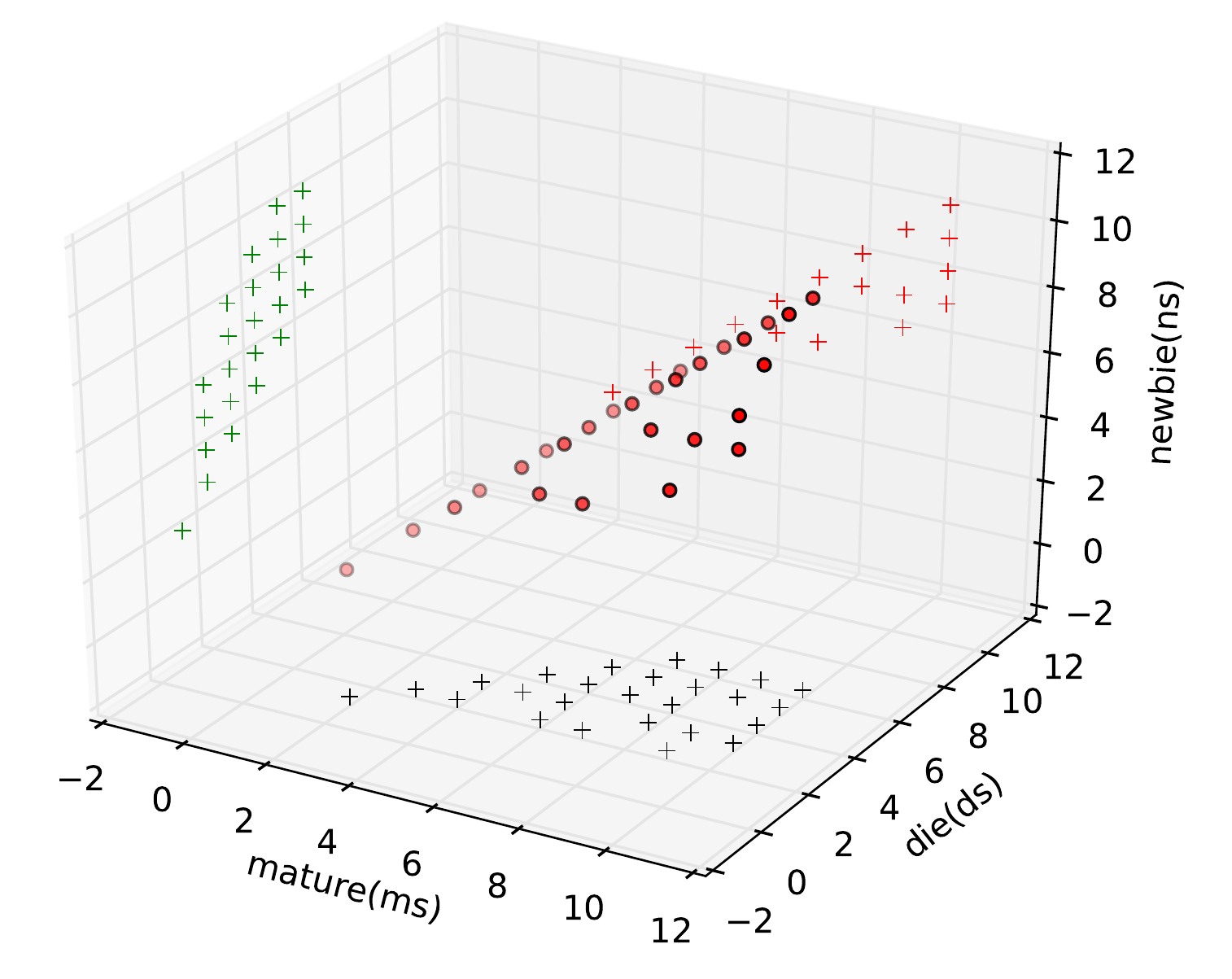}
\end{tabular}
\caption{Categories of the selfish and altruistic populations}
\label{fig:3}       
\end{figure}

It appeared convenient to distinguish 6 categories of the populations selected according to their results in the stress-test:
\begin{enumerate}
\item \textit{hopeless} quickly (before $10^4$ steps) fall down to less than 4 organisms without begetting any target genomes;
\item \textit{degenerating} beget one target genome and fall down to less than 4 organisms within $10^4$ steps;
\item \textit{dilatory} preserve genetic diversity, but do not beget even the first target genome in $10^4$ steps;
\item \textit{fallen} beget the first target genome before the step  $10^4$, but fall down to less than 4 organisms in the following evolution before the step $10^6$;
\item \textit{procrastinating} beget the first target genome before the step $10^4$ and preserve genetic diversity up to the step $10^6$, but do not beget all 8 target genomes in the given $10^6$ steps;
\item \textit{dexterous} succeeded in the stress-test.
\end{enumerate}

A population with the parameters $(ms,ds,ns)$ was assigned one of the above categories $x\in\overline{1,6}$ if 1) more than 50\% of the 100 experiments with these parameters resulted in the category $x$ or, if there is no such a category, 2) $x$ was the lowest category, which contained at least 16\% ($\approx$ 100\%/6 categories) of the results.

The outcome of the computational experiments is summarized in Fig. 4. Let me briefly comment on the distribution of the populations' parameters inside each of the above categories, after which we will pay more attention to the ``dexterous'' populations:
\begin{enumerate}[i]
\item most of the selfish populations remain ``hopeless''; however, the altruistic ``hopeless'' populations form a considerable but not dominant part of all the altruists;
\item all the selfish ``degenerating'' populations lie on the line projection where the newbie size is equal to the mature size (this is the maximum newbie size allowed by the prudent energy input constraint $ns\le ms\le 2ns$); the altruistic ``degenerating'' populations follow the same tendency, but with more dispersion, in addition they	agglomerate in the area with large mature size;
\item the ``dilatory'' selfish populations have a very low die size ($ds$), which on one hand gives them the chance to preserve the rich initial genetic diversity, but, on the other hand, deprives them of the chance to  transform under mutation-selection process towards the target genomes; the altruistic ``dilatory'' populations are subjects to the same shift, but they are more 	numerous and widespread in the parameter space;
\item  both altruistic and selfish ``fallen'' populations are rather an exception than a rule; the selfish ``fallen'' possess the die size equal to 1, which to all appearance is the minimal value needed to remain evolving and preserve a reasonable number of individuals simultaneously; the altruistic ``fallen'' have more scattered parameters;
\item the experiments revealed no ``procrastinating'' populations at all, which deepens the separation of the successful ``dexterous'' populations from all the other types;
\item there were no selfish populations among ``dexterous''; the altruistic ``dexterous'' populations meet the natural condition $ms>ds$ and lie mostly in the plane $ns=ms$, but there are some exceptions in case of low die size; there are more altruistic ``dexterous'' populations with high mature size, which (regardless of the evident explanation $ms>ds$) resembles the K-strategy of biological reproduction~(\cite{k-sel}).
\end{enumerate}


\begin{table}
\centering
\captionsetup{width=0.95\textwidth}
\caption{The results of the stress-test for the altruistic ``dexterous'' populations. Column ``steps'' corresponds to the rounded average number of steps taken to pass the test (to beget the 8-th target genome); column ``genomes'' shows the rounded average minimum number of different genomes in the population reached during the 100 tests}
\label{tab:1}       
\begin{tabular}{r||l}
\begin{tabular}{c|c|c|c|c}
\hline\noalign{\smallskip}
$ms$&$ds$&$ns$&steps&genomes\\
\noalign{\smallskip}\hline\noalign{\smallskip}
2&	1	&2&	130163	 & 85\\
3&	2	&3	&28947	 &38\\
4&	2	&4	&136373	& 73\\
4&	3	&4	&16133&	16\\
5&	3	&5	&32964	&47\\
5&	4	&5	&11347	&9\\
6&	2	&5	&140973	&68\\
6&	3	&6	&129216	&76\\
6&	4	&6	&17791&	17\\
6&	5	&6	&10562	&4\\
7&	2	&5	&64884	&15\\
7&	4	&7	&35189	&47\\
7&	5	&7	&13474	&6\\
7&	6	&7	&8998	&4\\
\end{tabular}&
\begin{tabular}{c|c|c|c|c}
\hline\noalign{\smallskip}
$ms$&$ds$&$ns$&steps&genomes\\
\noalign{\smallskip}\hline\noalign{\smallskip}
8&	3	&7	&100254	&64\\
8&	4	&8	&94569	&70\\
8&	5	&8	&20162	&23\\
8&	6	&8	&11375	&7\\
9&	2	&6	&156847	&7\\
9&	3	&7	&51527	&9\\
9&	5	&9	&40022	&35\\
9&	6	&9	&16814	&6\\
10&	3	&7&	171185&4\\
10	&3	&8	&129352	&44\\
10	&4	&9	&85892	&51\\
10	&5	&10&	76188	&62\\
10	&6&	10&	25370	&18\\
 & & & &
\end{tabular}
\end{tabular}\end{table}


\begin{figure*}
\centering
\includegraphics[width=0.8\textwidth]{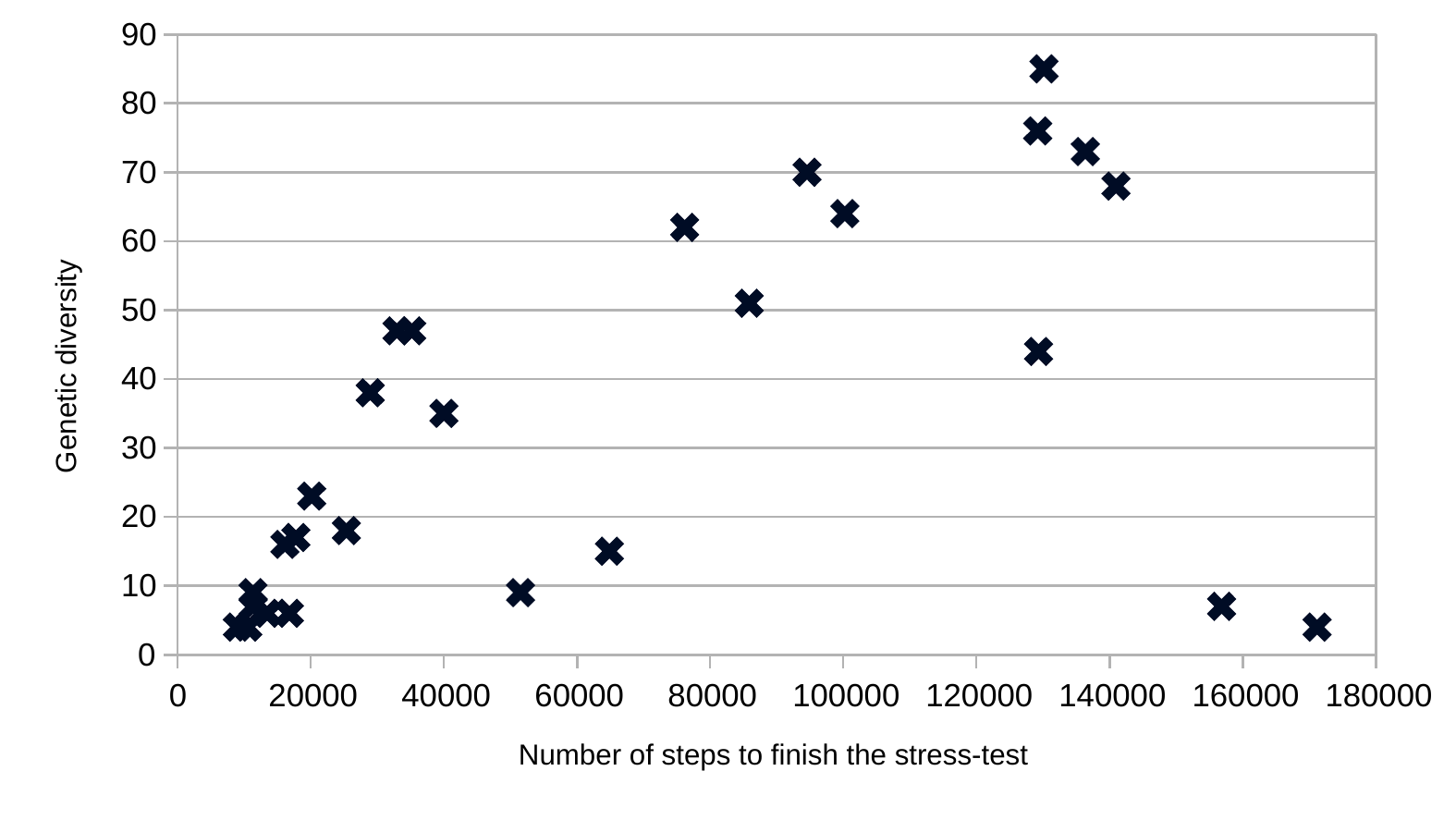}
\caption{The connection between ``speed'' and ``stability'' of evolution for the ``dexterous'' populations (based on Table \ref{tab:1})}
\label{fig:4}       
\end{figure*}

6-Altruistic graph of Fig. \ref{fig:4} shows that most of the ``dexterous'' populations lie in some 2D plane. The equation for this plane may be constructed by linear regression:
\begin{equation}
ms=-0.7ds+1.38ns+0.63,\quad R^2=0.9,\quad F=6\cdot 10^{-13}.
\end{equation}

The results of the stress-test with the ``dexterous'' altruistic populations  including the number of steps taken to beget all 8 target genomes and the minimum genetic variety during the evolution process are given in Tab. 1 and Fig. 5. 

\paragraph{Idyll-test.}
In the stress-test the populations tried to adapt to periodically changing environment.  The idyll-test puts the populations in the contrary situation: they live in an unchanging environment for a relatively long time, upon which the target genome changes drastically and a ``relaxed'' overspecialized population has to turn around to a new, totally different genetic purpose. Only the populations that proved to be ``dexterous'' in the stress-test (and those are all altruistic) were allowed for the idyll-test. 

In each idyll-test a population has to beget its first target genome $(0,0,0)$ in $10^4$ steps, following which the environment remains constant (the target genome is not switching) for either $10^4$ or $10^5$ steps. After such a ``peaceful age,'' the target genome changes to the opposite one, $(255,255,255)$, and the population has to beget the new target genome before the step $10^6$. The idyll-test experiments revealed only two categories of populations: ``fallen'' (see the description in the previous section) and ``succeeders'' (those who passed the idyll-test). 

Just as the stress-test, each idyll-test with the parameters $(ms,ds,ns)$ and under the constraints \eqref{constraints} was repeated 100 times. The result of such a series of experiments was similarly either the category that appeared in more than 50\% of the series or (if it does not exist) the lowest category having at least 16\% of the experiments.

\begin{table}
\centering
\caption{Results of the idyll-test}
\label{tab:2}       
\begin{tabular}{c||c||c}
\hline\noalign{\smallskip}
``dexterous'' & idyll lasts for $10^4$ & idyll lasts for $10^5$\\
\noalign{\smallskip}\hline\noalign{\smallskip}
27 & 
\begin{tabular}{c|c}
``fallen'' & ``succeeder''\\
2 & 25
\end{tabular}
&
\begin{tabular}{c|c}
``fallen'' & ``succeeder''\\
1 & 26
\end{tabular}
\end{tabular}
\end{table}

Almost all the altruistic ``dexterous'' populations that passed the idyll-test independently of the duration of the ``peaceful age.'' The ``fallen'' populations are as rare in the idyll-test as they were in the stress-test (see Tab. 2: two in the experiment with $10^4$ idyllic steps and one with $10^5$), which 
makes them look rather like an accidental bias than an informative result.

\section{Conclusion}

There are two hypotheses I would like to discuss in connection with the obtained results: one practical and one philosophical. The practical one concerns the evolutionary computation \cite{gen_alg_2,gen_alg_3}. Preservation of genetic diversity and ability to escape from overspecialization are very important features for any evolutionary computation approach. 
As we have seen, altruism-based evolutionary strategies may prove to be more adaptive and resistant to the common traps of evolution process, which means that altruism-based genetic algorithms are worth a further extensive study (already started in e.g. \cite{altru_gen2,altru_gen4,altru_gen3,altru_gen1}).


The second -- philosophical -- speculation is devoted to as much as the life in our Universe.
Considering the obtained experimental results, it is not unreasonable to assume that life with altruistic energy exchange is able to survive, adapt and develop in an unpredictable environment for longer periods compared to the selfish-based life. If we suppose that there are planets in the Universe, where life emerged in altruistic form, than group selection (at planetary level) will favor them as more resistant to the caprices of Nature. The life on such ``altruistic-biosphere'' planets should exist for longer periods and thus may become intelligent more often on average in comparison with the selfish life. The longer intelligent life exists -- the higher the chances for it to transform into a civilization of space travelers. Which  brings us to the final hypothesis: \textit{the highly developed extraterrestrial intelligent life we could encounter has more chances to be based on the altruistic principles} (see more on this topic in \cite{extra2,extra1}).

\bibliographystyle{ws-acs}
\bibliography{ivanko_refs}

\begin{thebibliography}{10}
\providecommand{\urlprefix}{}
\expandafter\ifx\csname urlstyle\endcsname\relax
  \providecommand{\doi}[1]{doi:\discretionary{}{}{}#1}\else
  \providecommand{\doi}{doi:\discretionary{}{}{}\begingroup
  \urlstyle{rm}\Url}\fi

\bibitem{altru6}
Axelrod, R., \emph{The evolution of cooperation} (Basic Books, New York, 1984).

\bibitem{gen_alg_2}
Back, T., \emph{Evolutionary algorithms in theory and practice: evolution
  strategies, evolutionary programming, genetic algorithms} (NY: Oxford
  University Press, New York, 1996).

\bibitem{games5}
Bomze, I.~M., Non-cooperative two-person games in biology: A classification,
  \emph{International Journal of Game Theory} \textbf{15} (1986) 31--57.

\bibitem{altru11}
Burtsev, M. and Turchin, P., Evolution of cooperative strategies from first
  principles, \emph{Nature} \textbf{440} (2006) 1041--1044.

\bibitem{altru9}
Cesta, A., Miceli, M., and Rizzo, P., Coexisting agents: Experiments on basic
  interaction attitude, \emph{Journal of Intelligent Systems} \textbf{11}
  (2011) 1--42.

\bibitem{recip2}
Clutton-Brock, T., Cooperation between non-kin in animal societies,
  \emph{Nature} \textbf{462} (2009) 51--57.

\bibitem{altru18}
Cooper, B. and Wallace, C., Group selection and the evolution of altruism,
  \emph{Oxford Economic Papers} \textbf{56} (2004) 307--330.

\bibitem{incl1}
Gardner, A. and West, S., Inclusive fitness: 50 years on, \emph{Philosophical
  Transactions of the Royal Society B: Biological Sciences} \textbf{369} (2014)
  1057--1062.

\bibitem{altru19}
Hamilton, W., The genetical evolution of social behaviour (1,2), \emph{Journal
  of Theoretical Biology} \textbf{7} (1964) 1--52.

\bibitem{games12}
Hofbauer, J. and Weibull, J.~W., Evolutionary selection against dominated
  strategies, \emph{Journal of Economic Theory} \textbf{71} (1996) 558--573.

\bibitem{extra2}
Impey, C., Spitz, A.~H., and Stoeger, W., \emph{Encountering Life in the
  Universe: Ethical Foundations and Social Implications of Astrobiology}
  (University of Arizona Press, 2013).

\bibitem{altru1}
Kropotkin, P., \emph{Mutual aid: a factor of evolution} (London, 1902).

\bibitem{radical}
Lehman, J. and Stanley, K.~O., Investigating biological assumptions through
  radical reimplementation, \emph{Artificial Life} \textbf{21} (2015) 21--46.

\bibitem{altru20}
Lehman, L. and Keller, L., The evolution of cooperation and altruism -- a
  general framework and a classification of models, \emph{Journal of
  Evolutionary Biology} \textbf{19} (2006) 1365--1376.

\bibitem{gen_alg_3}
Mitchell, M., \emph{An Introduction to Genetic Algorithms} (MIT Press,
  Cambridge, MA, USA, 1998).

\bibitem{altru_gen2}
Murata, T., Ishibuchi, H., and Gen, M., Specification of genetic search
  directions in cellular multi-objective genetic algorithms, in
  \emph{Proceedings of the First International Conference on Evolutionary
  Multi-Criterion Optimization}, EMO '01 (Springer-Verlag, London, UK, UK,
  2001), ISBN 3-540-41745-1, pp. 82--95.

\bibitem{bottle2}
Nei, M., Bottlenecks, genetic polymorphism and speciation, \emph{Genetics}
  \textbf{170} (2005) 1--4.

\bibitem{bottle1}
Nei, M., Maruyama, T., and Chakraborty, R., The bottleneck effect and genetic
  variability in populations, \emph{Evolution} \textbf{29} (1975) 1--10.

\bibitem{altru4}
Nowak, M.~A., Tarnita, C.~E., and Wilson, E.~O., The evolution of eusociality,
  \emph{Nature}  (2010) 1057--1062.

\bibitem{k-sel}
Parry, G.~D., The meanings of r- and k-selection, \emph{Oecologia} \textbf{48}
  (1981) 260--264.

\bibitem{altru_gen4}
Posp{\'i}chal, J. and Kvasni{\v{c}}ka, V., \emph{A Study of Altruism by Genetic
  Algorithm} (Springer London, London, 1999), pp. 507--520.

\bibitem{org}
Queller, D.~C. and Strassmann, J.~E., Beyond society: the evolution of
  organismality, \emph{Philosophical Transactions of the Royal Society of
  London B: Biological Sciences} \textbf{364} (2009) 3143--3155.

\bibitem{altru_gen3}
Ramteke, M. and Gupta, S.~K., Biomimicking altruistic behavior of honey bees in
  multi-objective genetic algorithm, \emph{Industrial \& Engineering Chemistry
  Research} \textbf{48} (2009) 9671--9685.

\bibitem{games1}
Smith, J., \emph{Evolution and the Theory of Games} (Cambridge University
  Press, 1982).

\bibitem{stasis}
Stenseth, N.~C. and Smith, J.~M., Coevolution in ecosystems: Red queen
  evolution or stasis?, \emph{Evolution} \textbf{38} (1984) 870--880.

\bibitem{recip1}
Trivers, R.~L., The evolution of reciprocal altruism, \emph{The Quarterly
  Review of Biology} \textbf{46} (1971) 35--57.

\bibitem{altru16}
Uyenoyama, M.~K., Evolution of altruism under group selection in large and
  small populations in fluctuating environments, \emph{Theoretical Population
  Biology} \textbf{15} (1979) 58--85.

\bibitem{extra1}
Vakoch, D.~A., \emph{Extraterrestrial Altruism Evolution and Ethics in the
  Cosmos} (Springer, 2014).

\bibitem{games15}
Weibull, J., \emph{Evolutionary Game Theory} (MIT Press, 1997).

\bibitem{altru_gen1}
Westerdale, T.~H., Altruism in the bucket brigade, in \emph{Proceedings of the
  Second International Conference on Genetic Algorithms on Genetic Algorithms
  and Their Application} (L. Erlbaum Associates Inc., Hillsdale, NJ, USA,
  1987), ISBN 0-8058-0158-8, pp. 22--26.

\bibitem{altru22}
Wu, B., Arranz, J., Du, J., Zhou, D., and Traulsen, A., Evolving synergetic
  interactions, \emph{Journal of The Royal Society Interface} \textbf{13}
  (2016).

\end{thebibliography}



\end{document}